\documentclass[ reprint, aps,]{revtex4-1}
\usepackage{amssymb}
\usepackage{amsmath}
\usepackage{graphicx}
\usepackage{dcolumn}
\usepackage{bm}

\begin{document}

\title{Quantum-enhanced SU(1,1) interferometry via a Fock state}
\author{Shuai Wang$^{1,\dagger }$, Jian-Dong Zhang$^{1}$, and Xuexiang Xu$%
^{2}$}

\address{$^1$ School of Mathematics and Physics, Jiangsu University of Technology,
Changzhou 213001, P.R. China, $^2$ Department of Physics, Jiangxi Normal University, Nanchang 330022, P.R. China,
\\$^{\dagger }$ Corresponding author: wshslxy@jsut.edu.cn}
\begin{abstract}
In this paper, we derive a general expression of the quantum Fisher
information of an SU(1,1) interferometer with an arbitrary state and a Fock
state as inputs by the phase-averaging method. Our results show that the
same quantum Fisher information can be obtained regardless of the specific
form of the arbitrary state. Then, we analytically prove that the parity
measurement can saturate the quantum Cramer-Rao bound when the estimated
phase sits at the optimal working point. For practical reasons, we
investigate the phase sensitivity when the arbitrary state is a coherent or
thermal state. We further show that a Fock state can indeed enhance the
phase sensitivity within a constraint on the total mean photon number inside
the interferometer.

\begin{description}
\item[PACS] 03.67.-a, 42.50.Dv
\end{description}
\end{abstract}

\maketitle

\section{Introduction}

Over the past decades, optical interferometers in the field of quantum
metrology have been widely used in the study of the phase estimation. A
Mach-Zehnder interferometer (MZI) is a basic and important tool to measure
an unknown phase, which typically contains two linear beam splitters. For an
MZI wih classical input states, the phase sensitivity of the phase
estimation is bounded by the shot-noise limit (SNL) $\Delta \phi =1/\sqrt{%
\bar{n}}$ \cite{1}, where $\bar{n}$ is the mean photon number inside the
interferometer. While, with nonclassical input states \cite%
{1,2,3,4,5,6,7,8,9,10}, the phase sensitivity can beat the SNL and even
reach the Heisenberg limit (HL) $\Delta \phi =1/\bar{n}$ \cite{11,12}.

The other way of beating the SNL is to use an interferometer in which the
mixing of the optical beams is done through a nonlinear transformation, such
as an SU (1,1) interferometer. The SU(1,1) was first proposed by Yurke et
al. \cite{13}, where the nonlinearity implemented by optical parametric
amplifiers (OPAs) or four-wave mixers. It has been shown that the SNL can be
beaten by an SU(1,1) interferometer in experiments with photons \cite{14}
and Bose--Einstein condensates \cite{15,16}. Recently, some theoretical work
with an SU(1,1) interferometer has also been done \cite%
{17,18,19,20,21,22,23,24}. For example, Plick et al. \cite{17} demonstrated
that an SU(1,1) interferometer with two coherent states as input states can
surpass the SNL by the intensity measurement. Li et al. analyzed the SU(1,1)
interferometer fed by coherent and squeezed vacuum states, and found that
the phase sensitivity can also reach the HL with balanced homodyne
measurement \cite{18} and parity measurement \cite{20}. With parity
measurement, Ma et al. \cite{24} studied the phase sensitivity of an SU(1,1)
interferometer with a thermal state and a squeezed vacuum state as inputs,
and found that the phase sensitivity can beat the SNL and approach the HL
with increasing input photon number.

In this manuscript, we consider the phase sensitivity of an SU(1,1)
interferometer with an arbitrary state $\hat{\rho}_{a}$ (for instance,
coherent or thermal states) and a Fock state $\left\vert n\right\rangle _{b}$
as inputs, i.e.,%
\begin{equation}
\hat{\rho}_{\text{in}}=\hat{\rho}_{a}\otimes \left\vert n\right\rangle
_{b}\left\langle n\right\vert ,  \label{1}
\end{equation}%
where an arbitrary state $\hat{\rho}_{a}$ can be written as $\hat{\rho}%
_{a}=\sum_{m,m^{\prime }}^{\infty }c_{m,m^{\prime }}\left\vert
m\right\rangle _{a}\left\langle m^{\prime }\right\vert $ in the basis of
Fock states. Such scheme uses a classical input mode, it is feasible to
create particle-entangled input states of large intensity, which are
essential ingredient in quantum metrology. As early as in 2013, Pezz\'{e}
and Smerzi \cite{25} has considered the phase sensitivity of an MZI with
such non-Gaussian states. They obtained the corresponding quantum Cram\'{e}%
r-Rao bound (QCRB) $\Delta \phi _{\text{QCRB}}=1/\sqrt{F_{Q}}=\left( 2n\text{%
Tr}\left[ \hat{\rho}_{a}\hat{a}^{\dag }\hat{a}\right] +n+\text{Tr}\left[
\hat{\rho}_{a}\hat{a}^{\dag }\hat{a}\right] \right) $, which provides a
sub-SNL phase sensitivity. Here, $F_{Q}$ is the quantum Fisher information.
Their results show that the same ultimate limit of the phase sensitivity can
be obtained for such input state in Eq. (\ref{1}). Subsequently, it has been
proved that the phase sensitivity with photon-number-resolving detection or
parity detection can saturate the QCRB \cite{25,26}. Following the work in
Ref. \cite{25}, we further investigate whether the above results are still
valid for an SU(1,1) interferometer. For practical reasons, we will explore
the improvement of the phase sensitivity induced by the single-mode Fock
state while the other input port is fed by a coherent or thermal state.

The remainder of this paper is organized as follows. In Sec. 2, we give the
quantum Fisher information of an SU (1,1) interferometer with the inputs
shown in Eq. (\ref{1}). In Sec. 3, we analytically prove that the phase
sensitivity via the parity measurement can saturate the QCRB. In Sec. 4, we
investigate the phase sensitivity with the mixing a coherent state (or
thermal state) and a Fock state. Finally, our conclusions are presented in
Sec. 5.

\section{Quantum Fisher information in an SU(1,1) interferometer}

In recent years, it has been shown without proper consideration of the
external phase reference that the quantum Fisher information (QFI) will give
different precision limits with different configurations of the unknown
phases for an MZI \cite{27,28} or an SU(1,1) interferometer \cite{29,30}.
Jarzyna and Demkowicz-Dobrza\'{n}ski \cite{27} first pointed out that the
naive calculation of QFI sometimes leads to overestimation of the limit.
They introduced the phase-averaging of the two-mode input state via a common
phase shift to rule out any additional phase reference that might give some
phase information to the measurement device. To do this, the possibility of
this overestimation is circumvented and the same QFI for different phase
configurations is obtained. Related to this, the fundamental limits of the
MZI and SU(1,1) interferometers when one of the inputs is a vacuum state
were discussed \cite{28,30}, respectively.

\begin{figure}[tbph]
\centering
\includegraphics[width=8cm]{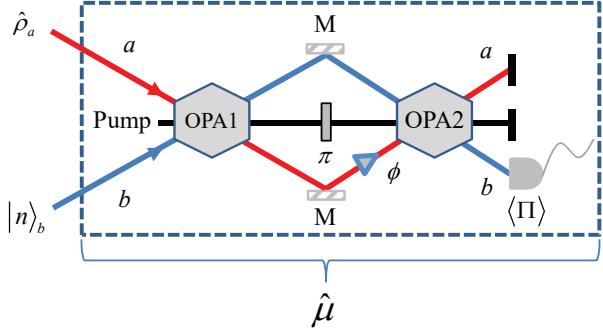}
\caption{(color online) Schematic diagram of an SU(1,1) interferometer. The
two input beams are in an arbitrary state and a Fock state, respectively.
OPA: optical parametric amplifier; $\hat{\Pi}$: parity measurement; M:
mirror.}
\end{figure}

Here, we consider the single-phase estimation of the SU(1,1) interferometer
with an arbitrary state $\hat{\rho}_{a}$ and a Fock state $\left\vert
n\right\rangle _{b}$ as the input state, as shown in Fig. 1. It is known
that the action of the OPA on a two-mode state is described by a two-mode
squeezing operator $\hat{S}_{2}\left( \xi \right) =\exp \left( \xi \hat{a}%
^{\dagger }\hat{b}^{\dagger }-\xi ^{\ast }\hat{a}\hat{b}\right) $ with $\xi
=ge^{i\theta }$, where $g$ and $\theta $ are the parametric gain and the
phase of the OPA, respectively. According to the work in Ref. \cite{27}, the
phase reference between the input state and the measurement can be removed
through the use of the phase-averaging operation, defined as%
\begin{eqnarray}
\Psi _{\text{avg}} &=&\int \frac{d\varphi }{2\pi }\hat{V}_{\varphi }^{A}\hat{%
V}_{\varphi }^{B}\left( \hat{\rho}_{a}\otimes \left\vert n\right\rangle
_{b}\left\langle n\right\vert \right) \hat{V}_{\varphi }^{A\dag }\hat{V}%
_{\varphi }^{B\dag }  \notag \\
&=&\sum_{m,m^{\prime }}^{\infty }\int \frac{d\varphi }{2\pi }e^{i\varphi
\left( m-m^{\prime }\right) }c_{m,m^{\prime }}\left\vert m\right\rangle
_{a}\left\langle m^{\prime }\right\vert \otimes \left\vert n\right\rangle
_{b}\left\langle n\right\vert  \notag \\
&=&\sum_{m=0}^{\infty }p_{m}\left\vert m\right\rangle _{a}\left\langle
m\right\vert \otimes \left\vert n\right\rangle _{b}\left\langle n\right\vert
,  \label{2}
\end{eqnarray}%
where $\hat{V}_{\varphi }^{A}=e^{i\varphi \hat{a}^{\dagger }\hat{a}}$, $\hat{%
V}_{\varphi }^{B}=e^{i\varphi \hat{b}^{\dagger }\hat{b}}$,$\,$and $%
p_{m}=c_{m,m}$ is a real positive number satisfying $\sum p_{m}=1$ \cite{31}%
. Obviously, the phase-averaging operation drops off all nondiagonal terms.
If the unknown phase \ to be estimated occurs in the upper or lower mode, it
is modeled by the unitary operator $\hat{U}_{\phi }^{u}=e^{i\hat{b}^{\dagger
}\hat{b}\phi }$ or $\hat{U}_{\phi }^{l}=e^{i\hat{a}^{\dagger }\hat{a}\phi }$%
. Upon leaving the estimated phase, the state evolves as%
\begin{eqnarray}
\Psi _{\text{avg}}^{\phi } &=&\hat{U}_{\phi }\hat{S}_{2}\left( \xi \right)
\Psi _{\text{avg}}\hat{S}_{2}^{\dag }\left( \xi \right) \hat{U}_{\phi
}^{\dag }  \notag \\
&=&\sum_{m=0}^{\infty }p_{m}\left\vert \psi _{m,n}\left( \phi \right)
\right\rangle _{ab}\left\langle \psi _{m,n}\left( \phi \right) \right\vert ,
\label{3}
\end{eqnarray}%
where%
\begin{eqnarray}
\left\vert \psi _{m,n}\left( \phi \right) \right\rangle &=&\hat{U}_{\phi }%
\hat{S}_{2}\left( \xi \right) \left\vert m\right\rangle _{a}\left\vert
n\right\rangle _{b}  \notag \\
&=&\sum_{k=0}^{\infty }c_{m,n,k}\left\vert m+k-l\right\rangle _{a}\left\vert
n+k-l\right\rangle _{b},  \label{4}
\end{eqnarray}%
with the coefficient $c_{m,n,k}$. When the phase shift $\phi $ occurs only
in the mode $a$ (the lower mode), $c_{m,n,k}$ reads as%
\begin{eqnarray}
c_{m,n,k} &=&\frac{e^{ik\theta }\tanh ^{k}g\sqrt{m!n!}}{k!\cosh ^{m+n+1}g}%
\sum_{l=0}^{\min [m,n]}\frac{e^{i\left( m+k-l\right) \phi -il\theta }\sinh
^{l}2g}{\left( -2\right) ^{l}}  \notag \\
&&\times \frac{\sqrt{\left( m+k-l\right) !\left( n+k-l\right) !}}{%
l!(m-l)!(n-l)!}.  \label{5}
\end{eqnarray}%
It can be seen that the coefficient $c_{m,n,k}$ is a function of the phase $%
\theta $ and the parametric gain $g$, as well as the phase shift $\phi $ to
be estimated.

By using the convexity of the QFI \cite{32,33} and noting that $\left\vert
\psi _{m,n}\left( \phi \right) \right\rangle $ and $\left\vert \psi
_{m^{\prime },n}\left( \phi \right) \right\rangle $ are orthogonal for $%
m\neq m^{\prime }$, we have%
\begin{equation}
F_{Q}\left( \Psi _{\text{avg}}^{\phi }\right) =\sum_{m=0}^{\infty
}p_{m}F_{Q}\left( \left\vert \psi _{m,n}\left( \phi \right) \right\rangle
\right) .  \label{6}
\end{equation}%
Especially, in the case of $n=0$, i.e., the above result reduces to that
results in Ref. \cite{30}. For pure states, the QFI of the $\left\vert \psi
_{m,n}\left( \phi \right) \right\rangle \,$\cite{34} is given by%
\begin{eqnarray}
&&F_{Q}\left( \left\vert \psi _{m,n}\left( \phi \right) \right\rangle \right)
\notag \\
&=&4\left( \left\langle \psi _{m,n}^{\prime }\left( \phi \right) \right\vert
\left. \psi _{m,n}^{\prime }\left( \phi \right) \right\rangle -\left\vert
\left\langle \psi _{m,n}^{\prime }\left( \phi \right) \right\vert \left.
\psi _{m,n}\left( \phi \right) \right\rangle \right\vert ^{2}\right) ,
\label{7}
\end{eqnarray}%
where $\left\vert \psi _{m,n}^{\prime }\left( \phi \right) \right\rangle
=\partial \left\vert \psi _{m,n}\left( \phi \right) \right\rangle /\partial
\phi $. For the convenience, we consider the phase shift in the mode $a$,
i.e., $\hat{U}_{\phi }=e^{i\hat{a}^{\dagger }\hat{a}\phi }$. Then$\,$, we
obtain%
\begin{equation}
F_{Q}\left( \left\vert \psi _{m,n}\left( \phi \right) \right\rangle \right)
=\left( 2mn+m+n+1\right) \sinh ^{2}\left( 2g\right) .  \label{8}
\end{equation}%
Substituting Eq. (\ref{8}) into Eq. (\ref{6}), we can immediately obtain the
QFI of the phase-averaged input state as follows%
\begin{equation}
F_{Q}\left( \Psi _{\text{avg}}^{\phi }\right) =\left( 2\bar{n}_{a}n+\bar{n}%
_{a}+n+1\right) \sinh ^{2}\left( 2g\right) ,  \label{9}
\end{equation}%
where $\bar{n}_{a}=$Tr$\left[ \hat{a}^{\dagger }\hat{a}\hat{\rho}_{a}\right]
=\sum_{m}mp_{m}$ is the average photon number of the arbitrary state $\hat{%
\rho}_{a}$. Obviously, compared with the case of an MZI considered in Ref.
\cite{25}, when one input of an SU(1,1) interferometer is a Fock state and
the other input leaves in an arbitrary state, one will also obtain the same
QFI.

Now we examine the effects of some parameters on the phase sensitivity. The
phase sensitivity of an interferometer is mainly determined by the total
mean photon number\ inside the interferometer. When an arbitrary state and a
Fock state are considered as the input state, the total mean photon number
inside the SU(1,1) interferometer is%
\begin{equation}
\bar{N}_{\text{Tot}}\equiv \left\langle \hat{a}_{1}^{\dag }\hat{a}_{1}+\hat{b%
}_{1}^{\dag }\hat{b}_{1}\right\rangle =\left( \bar{n}_{a}+n\right) \cosh
2g+2\sinh ^{2}g,  \label{10}
\end{equation}%
where we have used the relationship between the operators of the out modes ($%
\hat{a}_{1},\hat{b}_{1}$) and the input modes ($\hat{a},\hat{b}$) for an OPA
\begin{eqnarray}
\hat{a}_{1} &=&\hat{a}\cosh g-e^{i\theta }\hat{b}^{\dagger }\sinh g,  \notag
\\
\hat{b}_{1} &=&-e^{i\theta }\hat{a}^{\dagger }\sinh g+\hat{b}\cosh g.
\label{11}
\end{eqnarray}%
The ultimate bounds of the phase sensitivity with a general input state $%
\hat{\rho}_{\text{in}}$ is given by the quantum Cram\'{e}r-Rao bound (QCRB)
\cite{34,35}, $\Delta \phi _{\text{QCRB}}=1/\sqrt{F_{Q}}$. For the input
state expressed in Eq. (\ref{1}), the corresponding QCRB of the SU(1,1)
interferometer reads
\begin{equation}
\Delta \phi _{\text{QCRB}}=\frac{1}{\sqrt{\left( 2\bar{n}_{a}n+\bar{n}%
_{a}+n+1\right) \sinh ^{2}\left( 2g\right) }}.  \label{12}
\end{equation}

In Fig. 2, we plot that the QCRB varies with the total mean photon number
inside the interferometer for different values of the parameters $\bar{n}%
_{a} $, $n$ and $g$. From Fig. 2 (a), we can see that the phase sensitivity
greatly improves with increasing $n$. For given the parameters $n$ and $g$,
the phase sensitivity can not approach the HL with increasing the total mean
photon number. While for given the parameters $\bar{n}_{a}$ and some values
of $n$, the phase sensitivity will approach the HL with increasing the total
mean photon number determined by the parametric gain $g$ as shown in Fig. 2
(b). It can be seen that the Fock state can greatly improve the phase
sensitivity within a constraint on the total mean photon number inside the
SU(1,1) interferometer, especially in the case of single-photon Fock state.

\begin{figure}[tbph]
\centering\includegraphics[width=8cm]{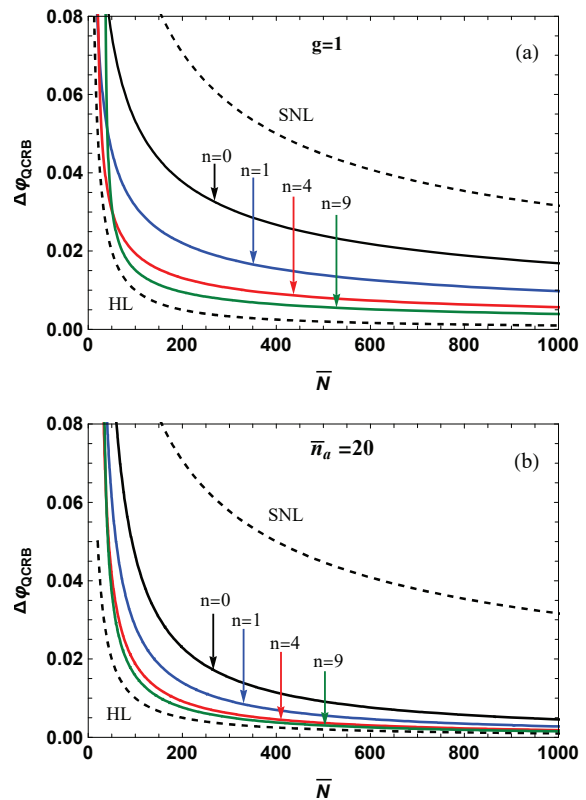}
\caption{(color online) The QCRB as a function of the total mean photon
number inside the SU(1,1) interferometer. (a) Only the mean photon number $%
\bar{n}_{a}$ of an arbitrary state is being changed; (b) Only the parametric
gain $g$ is being changed. The upper black dashed line denotes the SNL,
while the below black dashed line represents the HL.}
\end{figure}

\section{Parity measurement for an arbitrary state and a Fock state}

So far, we have investigate the QCRB. It is well known that this limit can
be saturated by the optimal generalized measurement \cite{34,35}. Here, we
consider the parity measurement and demonstrate that it saturates the QCRB
and is optimal for our considered scheme. The parity operator is given by $%
\hat{\Pi}=\exp \left[ i\pi \hat{a}^{\dagger }\hat{a}\right] $ or $\exp \left[
i\pi \hat{b}^{\dagger }\hat{b}\right] $, which distinguishes even or odd
numbers of photons. The parity measurement is actually to obtain the
expectation value of such observable viable. In the traditional operator
methods, the parity signal on one output mode of an interferometer reads $%
\left\langle \hat{\Pi}\left( \varphi \right) \right\rangle =$Tr$\left[ \hat{%
\rho}_{\text{out}}\hat{\Pi}\right] $. In pricinple, it is very difficult to
directly calculate the parity signal in an interferometer by such operator
method, especially for an SU(1,1) interferometer. Therefore, the expectation
value of the parity operator is usually obtained by the value of Wigner
function at the origin, i.e., $\left\langle \hat{\Pi}\left( \varphi \right)
\right\rangle =\pi W\left( \alpha ,0\right) $ (or $\pi W\left( 0,\beta
\right) $), where $W\left( \alpha ,\beta \right) $ is the Wigner function of
the output state of an MZI or SU(1,1) interferometer. However, it is still
difficult to derive the parity signal by the transformation of Wigner
functions when non-Gaussian states are considered as the input state,
especially for an SU(1,1) interferometer.

In our previous work, we present an alternative operator method \cite{36} in
the Heisenberg representation to analyze the signal of the parity
measurement within an SU(1,1) interferometer, i.e.,\textbf{\ }
\begin{equation}
\left\langle \hat{\Pi}\left( \phi \right) \right\rangle =\text{Tr}\left[
\hat{\rho}_{\text{in}}\hat{\mu}\left( \xi ,\phi \right) \right] ,  \label{13}
\end{equation}%
where the measurement operator $\hat{\mu}\left( \xi ,\phi \right) $ (a
Hermitian operator) completely describes the whole operation of the parity
measurement combined with an SU(1,1) interferometer. Consequently, one can
directly investigate the parity signal in terms of the input states. When
the estimated phase occurs in the upper or lower mode of the SU(1,1)
interferometer, we can obtain the same measurement operator $\hat{\mu}\left(
\xi ,\phi \right) $. Its normal ordered form is%
\begin{eqnarray}
\hat{\mu}\left( \xi ,\phi \right) &=&\frac{1}{1+2\sin ^{2}\frac{\phi }{2}%
\sinh ^{2}2g}\exp \left[ \hat{a}^{\dagger }\hat{b}^{\dagger }M^{\ast }\right]
\label{14} \\
&&\colon \exp \left[ -\hat{a}^{\dagger }\hat{a}C-\hat{b}^{\dagger }\hat{b}D%
\right] \colon \exp \left[ \hat{a}\hat{b}M\right] ,  \notag
\end{eqnarray}%
where%
\begin{equation}
M=\frac{e^{-i\theta }\left( i\sin \phi -2\sin ^{2}\frac{\phi }{2}\cosh
2g\right) \sinh 2g}{1+2\sin ^{2}\frac{\phi }{2}\sinh ^{2}2g},  \label{15}
\end{equation}%
\begin{equation}
C=\frac{2\sin ^{2}\frac{\phi }{2}\sinh ^{2}2g}{1+2\sin ^{2}\frac{\phi }{2}%
\sinh ^{2}2g},D=\frac{2+2\sin ^{2}\frac{\phi }{2}\sinh ^{2}2g}{1+2\sin ^{2}%
\frac{\phi }{2}\sinh ^{2}2g},  \label{16}
\end{equation}%
with the relation $CD=\left\vert M\right\vert ^{2}$. Noting that the
eigenvalue equations of the annihilation operator $\hat{a}\left\vert \alpha
\right\rangle =\alpha \left\vert \alpha \right\rangle \,$\ and the
properties of the normal ordered form of the operators \cite{37,38}, one can
directly obtain the parity signal when a two-mode coherent state $\hat{\rho}%
_{\text{in}}=\left\vert \alpha \right\rangle _{a}\left\vert \beta
\right\rangle _{b}\left. _{b}\left\langle \beta \right\vert \right. \left.
_{a}\left\langle \alpha \right\vert \right. $ is fed into an SU (1,1)
interferometer, i.e.,
\begin{eqnarray}
\left\langle \hat{\Pi}\left( \phi \right) \right\rangle &=&\frac{1}{1+2\sin
^{2}\frac{\phi }{2}\sinh ^{2}2g}  \notag \\
&&\exp \left[ 2\mathrm{Re}(\alpha \beta M)-\left\vert \alpha \right\vert
^{2}C-\left\vert \beta \right\vert ^{2}D\right] .  \label{17}
\end{eqnarray}%
Therefore, based on our method, it is convenient to derive the signal of the
parity measurement when the input state is expressed in the coherent state
basis.

\subsection{Parity measurement as the detection strategy}

For our purpose, we first consider the interferometer input state with
asymmetric two-mode Fock states with arbitrary photon numbers, i.e., $%
\left\vert \text{in}\right\rangle =\left\vert m\right\rangle _{a}\left\vert
n\right\rangle _{b}=\left( \hat{a}^{\dagger m}\hat{b}^{\dagger n}\right) /%
\sqrt{m!n!}\left\vert 0\right\rangle _{a}\left\vert 0\right\rangle _{b}$.
For the convenience, it is useful to expand the Fock state in the coherent
state basis,%
\begin{equation}
\left\vert k\right\rangle =\frac{\partial ^{k}}{\sqrt{k!}\partial x^{k}}\int
\frac{d^{2}\alpha }{\pi }e^{-\frac{1}{2}\left\vert \alpha \right\vert
^{2}+x\alpha ^{\ast }}\left\vert \alpha \right\rangle |_{x=0}.  \label{18}
\end{equation}%
According to Eqs. (\ref{13}) and (\ref{14}), when an asymmetric two-mode
Fock state is fed into an SU(1,1) interferometer, we can obtain the parity
signal as following%
\begin{eqnarray}
&&\left\langle \hat{\Pi}_{b}\left( \phi \right) \right\rangle _{m,n}  \notag
\\
&=&\frac{1}{1+2\sin ^{2}\frac{\phi }{2}\sinh ^{2}2g}\frac{\partial ^{2m+2n}}{%
m!n!\partial x^{m}\partial y^{n}\partial q^{m}\partial p^{n}}  \notag \\
&&\exp \left[ xyM^{\ast }+pqM+xq\left( 1-C\right) \right.  \notag \\
&&\left. +py\left( 1-D\right) \right] |_{x,y,p,q=0},  \label{19}
\end{eqnarray}%
where we have used the integral formula \cite{39}
\begin{equation}
\int \frac{d^{2}z}{\pi }e^{\zeta \left\vert z\right\vert ^{2}+\xi z+\eta
z^{\ast }}=-\frac{1}{\zeta }e^{-\frac{\xi \eta }{\zeta }},  \label{a}
\end{equation}%
whose convergent condition is Re$\left( \zeta \right) <0$. Equation (\ref{19}%
) can be further rewritten as%
\begin{eqnarray}
&&\left\langle \hat{\Pi}_{b}\left( \phi \right) \right\rangle _{m,n}  \notag
\\
&=&m!n!\left( -1\right) ^{n}\left( 1+2\sin ^{2}\frac{\phi }{2}\sinh
^{2}2g\right) ^{-\left( m+n+1\right) }  \notag \\
&&\times \sum_{k=0}^{\min [m,n]}\frac{\left[ \left( \sin ^{2}\phi +4\sin ^{4}%
\frac{\phi }{2}\cosh ^{2}2g\right) \sinh ^{2}2g\right] ^{k}}{\left(
-1\right) ^{k}\left( k!\right) ^{2}\left( m-k\right) !\left( n-k\right) !}.
\label{20}
\end{eqnarray}%
Specially, in the limit $\phi \rightarrow 0$, by expanding $\left\langle
\hat{\Pi}_{b}\left( \phi \right) \right\rangle _{m,n}$ in the Taylor series
around $\phi =0$, we obtain%
\begin{eqnarray}
&&\left\langle \hat{\Pi}_{b}\left( \phi \right) \right\rangle _{m,n}|_{\phi
\rightarrow 0}  \notag \\
&=&\left( -1\right) ^{n}\left[ 1-\frac{2mn+m+n+1}{2!}\phi ^{2}\sinh
^{2}\left( 2g\right) \right.  \notag \\
&&\left. +O\left( \phi ^{4}\right) \right] .  \label{21}
\end{eqnarray}%
Therefore, for an interferometer using an arbitrary state $\hat{\rho}%
_{a}=\sum_{m,m^{\prime }}c_{m,m^{\prime }}\left\vert m\right\rangle
\left\langle m^{\prime }\right\vert $ in mode $a$ and a Fock state $%
\left\vert n\right\rangle \left\langle n\right\vert $ in mode $b$, the
corresponding signal of the parity measurement can be immediately obtained%
\begin{equation}
\left\langle \hat{\Pi}_{b}\left( \phi \right) \right\rangle _{\hat{\rho}%
_{a},n}=\sum_{m=0}^{\infty }p_{m}\left\langle \hat{\Pi}_{b}\left( \phi
\right) \right\rangle _{m,n}.  \label{22}
\end{equation}%
It should be pionted out that, in the case of $m\neq m^{\prime }$, one can
prove that $\left\langle \hat{\Pi}_{b}\left( \phi \right) \right\rangle
_{m,m^{\prime },n}=$Tr$\left[ \left\vert m\right\rangle _{a}\left\langle
m^{\prime }\right\vert \left\vert n\right\rangle _{b}\left\langle
n\right\vert \hat{\mu}\left( \xi ,\phi \right) \right] =0$. Therefore, all
nondiagonal terms of the arbitrary state $\hat{\rho}_{a}$ do not contribute
to the parity signal. As mentioned in the above, for calculating the QFI,
these nondiagonal terms of $\hat{\rho}_{a}$ are also ruled out by the
phase-averaging operation. In the limit of $\phi \rightarrow 0$, combining
Eqs. (\ref{21}) and (\ref{22}) we can directly obtain

\begin{eqnarray}
&&\left\langle \hat{\Pi}_{b}\left( \phi \right) \right\rangle _{\hat{\rho}%
_{a},n}  \notag \\
&=&\left( -1\right) ^{n}\left[ 1-\frac{2\bar{n}_{a}n+\bar{n}_{a}+n+1}{2!}%
\sinh ^{2}\left( 2g\right) \phi ^{2}\right.  \notag \\
&&\left. +O\left( \phi ^{4}\right) \right] ,  \label{23}
\end{eqnarray}%
where we have used the mean photon number $\bar{n}_{a}$ of the state $\hat{%
\rho}_{a}$ and the normalizing condition Tr$\left[ \hat{\rho}_{a}\right] =1$%
. Therefore, when the estimated phase approaches to zero (the optimal
working point), our results indicate that the same parity signal will be
obtained for given the same mean photon number of the arbitrary state $\hat{%
\rho}_{a}$. When the estimated phase somewhat deviates from zero, different
inputs may give different phase resolutions, as shown in the following.

According to the error propagation theorem, the phase sensitivity is
obtained by

\begin{equation}
\Delta \phi =\frac{\left( 1-\left\langle \hat{\Pi}_{b}\left( \phi \right)
\right\rangle ^{2}\right) ^{1/2}}{\left\vert \frac{\partial \left\langle
\hat{\Pi}_{b}\left( \phi \right) \right\rangle }{\partial \phi }\right\vert }%
.  \label{24}
\end{equation}%
Substituting Eq. (\ref{22}) into Eq. (\ref{24}), we can analytically
demonstrate that the parity measurement saturates the QCRB and is the
locally optimal for our scheme.

\subsection{ Phase sensitivity with single-mode Fock state}

In this section, we consider the arbitrary state $\hat{\rho}_{a}$ to be a
coherent state or a thermal state. Additionall, we investigate the
corresponding phase resolutions and phase sensitivity with leaving the other
input in a Fock state. First, we consider a coherent state and a Fock state $%
\hat{\rho}_{\text{in}}=\left\vert \alpha \right\rangle _{a}\left\vert
n\right\rangle _{b}\left. _{b}\left\langle n\right\vert \right. \left.
_{a}\left\langle \alpha \right\vert \right. $ as the input state of an
SU(1,1) interferometer. The Fock state can be expressed in the coherent
state basis as shown in Eq. (\ref{18}). Then, according to Eqs. (\ref{13})
and (\ref{14}), the signal of parity measurement is%
\begin{eqnarray}
&&\left\langle \hat{\Pi}_{b}\left( \phi \right) \right\rangle _{\alpha ,n}
\notag \\
&=&\text{Tr}\left[ \left\vert \alpha \right\rangle _{a}\left\vert
n\right\rangle _{b}\left. _{b}\left\langle n\right\vert \right. \left.
_{a}\left\langle \alpha \right\vert \right. \hat{\mu}\left( \xi ,\phi
\right) \right]  \notag \\
&=&\frac{\exp \left( -\left\vert \alpha \right\vert ^{2}C\right) }{1+2\sin
^{2}\frac{\phi }{2}\sinh ^{2}2g}\frac{\partial ^{2n}}{n!\partial
x^{n}\partial y^{n}}  \notag \\
&&\exp \left[ xy\left( 1-D\right) +x\alpha M^{\ast }+y\alpha ^{\ast }M\right]
_{x,y=0},  \label{25}
\end{eqnarray}%
or

\begin{equation}
\left\langle \hat{\Pi}_{b}\left( \phi \right) \right\rangle =\frac{\left(
1-D\right) ^{n}\exp \left( -\left\vert \alpha \right\vert ^{2}C\right) }{%
1+2\sin ^{2}\frac{\phi }{2}\sinh ^{2}2g}L_{n}\left( \frac{\left\vert \alpha
\right\vert ^{2}\left\vert M\right\vert ^{2}}{D-1}\right) ,  \label{26}
\end{equation}%
where $L_{n}\left( x\right) $ is the Laguerre polynomials \cite{40}.

We turn to consider the arbitrary state $\hat{\rho}_{a}$ to be a thermal
state $\hat{\rho}_{\text{th}}$. For the convenience, the thermal state $\hat{%
\rho}_{\text{th}}$ is expressed in $P$-representation, i.e.,%
\begin{equation}
\rho _{\text{th}}=\frac{1}{\bar{n}_{\text{th}}}\int \frac{d^{2}\alpha }{\pi }%
\exp \left( -\frac{1}{\bar{n}_{\text{th}}}\left\vert \alpha \right\vert
^{2}\right) \left\vert \alpha \right\rangle _{a}\left\langle \alpha
\right\vert .  \label{27}
\end{equation}%
When a thermal state and a Fock state $\hat{\rho}_{\text{in}}=\hat{\rho}_{%
\text{th}}\otimes \left\vert n\right\rangle _{bb}\left\langle n\right\vert $
are injected into the SU(1,1) interferometer, the corresponding signal of
parity measurement is%
\begin{eqnarray}
&&\left\langle \hat{\Pi}_{b}\left( \phi \right) \right\rangle _{\rho _{\text{%
th}},n}  \notag \\
&=&\frac{1}{\bar{n}_{\text{th}}}\int \frac{d^{2}\alpha }{\pi }e^{-\frac{1}{%
\bar{n}_{\text{th}}}\left\vert \alpha \right\vert ^{2}}\text{Tr}\left[
\left\vert \alpha \right\rangle _{a}\left\vert n\right\rangle _{b}\left.
_{b}\left\langle n\right\vert \right. \left. _{a}\left\langle \alpha
\right\vert \right. \hat{\mu}\left( \xi ,\phi \right) \right]  \notag \\
&=&\frac{\left( 1+\bar{n}_{\text{th}}C-D\right) ^{n}\left( 1+\bar{n}_{\text{%
th}}C\right) ^{-n-1}}{1+2\sin ^{2}\frac{\phi }{2}\sinh ^{2}2g},  \label{28}
\end{eqnarray}%
where we have used Eq. (\ref{25}).

Within the constraint of the same total mean photon number inside the
interferometer, in Fig. 3 we plot the $\left\langle \hat{\Pi}_{b}\left( \phi
\right) \right\rangle $ against the estimated phase $\phi $ for some values
of the mean photon number $\bar{n}_{a}=\left\vert \alpha \right\vert ^{2}$ ($%
\bar{n}_{a}=\bar{n}_{\text{th}}$) and $n$ of the input state $\hat{\rho}%
_{a}\otimes \left\vert n\right\rangle _{b}\left\langle n\right\vert $. From
Fig. 3, one can see that, in the limit of $\phi \rightarrow 0$, the value of
$\left\langle \hat{\Pi}_{b}\left( \phi \right) \right\rangle _{\alpha ,n}$
is almost exactly equal to that of $\left\langle \hat{\Pi}_{b}\left( \phi
\right) \right\rangle _{\rho _{\text{th}},n}$, which is consistent with Eq. (%
\ref{23}). In addition, the central peak of the $\left\langle \hat{\Pi}%
_{b}\left( \phi \right) \right\rangle $ in the limit of $\phi \rightarrow 0$
narrows as $n$ increases, which indicates that the single-mode Fock state
can further improve the phase resolution. Compared with the thermal and Fock
inputs, the coherent and Fock inputs can give better phase resolution.

\begin{figure}[tbph]
\centering\includegraphics[width=8cm]{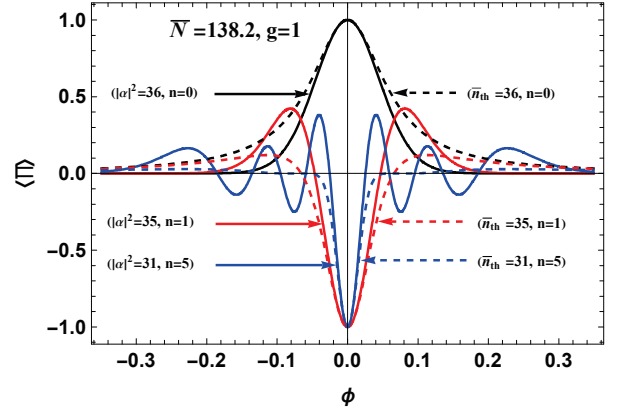}
\caption{(color online) Within the constrain of the total mean photon number
inside the SU (1,1) interferometer, the parity signal versus the estimated
phase $\protect\phi $ with a coherent state (or a thermal state) and a Fock
state QCRB as input states.}
\end{figure}

Finally, we investigate the effects of Fock states on the phase sensitivity.
In Fig. 4, we draw the phase sensitivity as a function of the estimated
phase $\phi $ for the same total mean photon number inside the
interferometer. From Fig. 4, the Fock state indeed can improve the phase
sensitivity, even with a single-photon state. Under the constraint of the
same total mean photon number inside the interferometer, the coherent state
and Fock state also gives better phase sensitivity than the mixing of a
thermal state and a Fock state when the estimated phase $\phi $ slightly
deviates from zero as shown in Fig. 4. In the limit of $\phi \rightarrow 0$,
we have proved that the parity measurement saturates the QCRB, thus the
variation of the phase sensitivity with the total mean photon number inside
the interferometer is same to Fig. 2.

\begin{figure}[tbph]
\centering\includegraphics[width=8cm]{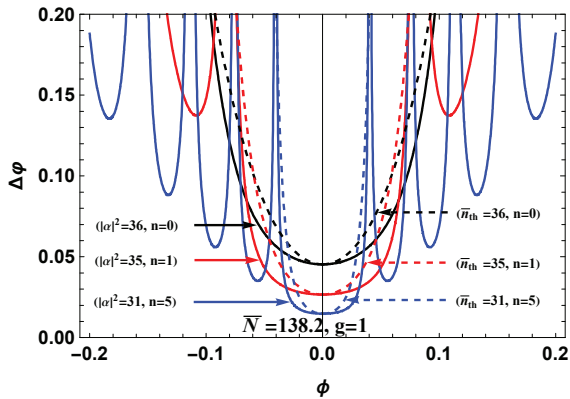}
\caption{(color online) The phase sensitivity as a function of the estimated
phaset $\protect\phi $ with a coherent state (or a thermal state) and a Fock
state as input states. The solid lines represent the product of a coherent
state and a Fock state,while the dashed lines represent the mixing of a
thermal state and a Fock state.}
\end{figure}

\section{Conclusions}

In summary, we showed that the same QCRB can be obtained when a single-mode
Fock state is injected into one of two ports of an SU(1,1) interferometer,
while the other port is fed by an arbitrary state. Through a new method, we
analytically demonstrated that the parity measurement can saturate the QCRB
and is optimal for our scheme in the limit of $\phi \rightarrow 0$.
Subsequently, we considered the arbitrary state to be a coherent state and a
thermal state, respectively. Then, we investigated the phase sensitivity
obtained by the parity measurement. Our results show that, within the
constraint of the same mean photon number inside the interferometer, a
single-mode Fock state can further improve the phase resolution and the
phase sensitivity when compared with only a coherent or thermal state as an
interferometer state.

\section*{Acknowledgments}

This work is supported by the National Natural Science Foundation of China
(Grant No. 11665013), and the Natural Science Foundation of Jiangsu Province
of China (Grant No. BK20140253).



\begin{thebibliography}{99}
\bibitem{1} C. M. Caves, Quantum-mechanical noise in an interferometer,
Phys. Rev. D \textbf{23}, 1693 (1981).

\bibitem{2} R. A. Campos, C. C. Gerry, and A. Benmoussa, Optical
interferometry at the Heisenberg limit with twin Fock states and parity
measurements, Phys. Rev. A \textbf{68}, 023810 (2003).

\bibitem{3} J. P. Dowling, Quantum optical metrology-the lowdown on
high-NOON states, Contemp. Phys. \textbf{49}, 125 (2008).

\bibitem{4} P. M. Anisimov, G. M. Raterman, A. Chiruvelli, W. N. Plick, S.
S. Huver, H. Lee, and J. P. Dowling, Quantum Metrology with Two-Mode
Squeezed Vacuum: Parity Detection Beats the Heisenberg Limit, Phys. Rev.
Lett. \textbf{104}, 103602 (2010).

\bibitem{5} J. Joo, W. J. Munro, T. P. Spiller, Quantum Metrology with
Entangled Coherent States, Phys. Rev. Lett. \textbf{107}, 083601 (2011).

\bibitem{6} S. Y. Lee, C. W. Lee, J. Lee, and H. Nha, Quantum phase
estimation using path-symmetric entangled states, Sci. Rep. \textbf{6},
30306 (2016).

\bibitem{7} Y. Ouyang, S. Wang, and L. Zhang, Quantum optical interferometry
via the photon added two-mode squeezed vacuum states, J. Opt. Soc. Am. B
\textbf{33}, 1373 (2016).

\bibitem{8} J. H. Xu, J. Z. Wang, A. X. Chen, Y. Li, and G. R. Jin, Optimal
phase estimation with photon-number difference measurement using twin-Fock
states of light, Chin. Phys. B \textbf{28}, 120303 (2019).

\bibitem{9} S. Wang, X. X. Xu, Y. J. Xu, and L. J. Zhang, Quantum
interferometry via a coherent state mixed with a photon-added squeezed
vacuum state, Opt. Commum. \textbf{444}, 102 (2019).

\bibitem{10} L. L Hou, S. Wang, and X. F. Xu, Optical enhanced
interferometry with two-mode squeezed twin-Fock states and parity detection,
Chin. Phys. B \textbf{29}, 034203 (2020).

\bibitem{11} S. L. Braunstein, Quantum Limits on Precision Measurements of
Phase, Phys. Rev. Lett. \textbf{69}, 3598 (1992).

\bibitem{12} M. J. Holland and K. Burnett, Interferometric detection of
optical phase shifts at the Heisenberg limit, Phys. Rev. Lett. \textbf{71},
1355 (1993).

\bibitem{13} B. Yurke, S. L. McCall, and J. R. Klauder, SU(2) and SU(1, 1)
interferometers, Phys. Rev. A \textbf{33}, 4033 (1986).

\bibitem{14} F. Hudelist, J. Kong, C. Liu, J. Jing, Z. Y. Ou, and W. Zhang,
Quantum metrology with parametric amplifier-based photon correlation
interferometers,\ Nat. Commun. \textbf{5}, 3049 (2014).

\bibitem{15} J. Peise, I. Kruse, K. Lange, B. L\"{u}ke, L. Pezz\`{e}, J.
Arlt, W. Ertmer, K. Hammerer, L. Santos, A. Smerzi, and C. Klempt,
Satisfying the Einstein--Podolsky--Rosen criterion with massive particles,\
Nat. Commun. \textbf{6}, 8984 (2015).

\bibitem{16} D. Linnemann, H. Strobel, W. Muessel, J. Schulz, R. J.
Lewis-Swan, K. V. Kheruntsyan, and M. K. Oberthaler, Quantum-enhanced
sensing based on time reversal of nonlinear dynamics,\ Phys. Rev. Lett.
\textbf{117}, 013001 (2016).

\bibitem{17} W. N. Plick, J. P. Dowling, and G. S. Agarwal,
coherent-light-boosted, sub-shot noise, quantum interferometry, New J. Phys.
\textbf{12}, 083014 (2010)

\bibitem{18} D. Li, C. H. Yuan, Z. Y. Ou, and W. Zhang, The phase
sensitivity of an SU(1,1) interferometer with coheren and squeezed-vacuum
light, New J. Phys. \textbf{16}, 073020 (2014)

\bibitem{19} M. Gabbrielli, L. Pezz\`{e}, and A. Smerzi, Spin-mixing
interferometry with Bose--Einstein condensates, Phys. Rev. Lett. \textbf{115}%
, 163002 (2015).

\bibitem{20} D. Li, B. T. Gard, Y. Gao, C. H. Yuan, W. Zhang, H. Lee, and J.
P. Dowling, Phase sensitivity at the Heisenberg limit in an SU(1, 1)
interferometer via parity detection,\ Phys. Rev. A \textbf{94}, 063840
(2016).

\bibitem{21} S. S. Szigeti, R. J. Lewis-Swan, and S. A. Haine, Pumped-up
SU(1, 1) interferometry, Phys. Rev. Lett. \textbf{118}, 150401 (2017).

\bibitem{22} Q. K. Gong, X. L. Hu, D. Li, C. H. Yuan, Z. Y. Ou, and W. P.
Zhang, Intramode-correlation-enhanced phase sensitivities in an SU (1, 1)
interferometer,\ Phys. Rev. A \textbf{96}, 033809 (2017).

\bibitem{23} L. L. Guo, Y. F. Yu, and Z. M. Zhang, Improving the phase
sensitivity of an SU(1,1) interferometer with photon-added squeezed vacuum
light, Opt. Express \textbf{28}, 29099 (2018).

\bibitem{24} X. P. Ma, C. L. You, S. Adhikari, E. S. Matekole, R. T.
Glasser, H. Lee, and J. P. Dowling. Sub-shot-noise-limited phase estimation
via SU(1,1) interferometer with thermal states, Opt. Express \textbf{26},
18492 (2018).

\bibitem{25} L. Pezz\`{e} and A. Smerzi, Ultrasensitive two-mode
interferometry with single-mode number squeezing, Phys. Rev. Lett. \textbf{%
110}, 163604 (2013).

\bibitem{26} S. Wang, Y. Wang, L. Zhai, and L. Zhang, Two-mode quantum
interferometry with single-mode Fock state and pariy detection, J. Opt. Soc.
Am. B \textbf{35}, 1046 (2018).

\bibitem{27} M. Jarzyna, R. Demkowicz-Dobrza\'{n}ski, Quantum interferometry
with and without an external phase reference, Phys. Rev. A \textbf{85}, 3353
(2012).

\bibitem{28} M. Takeoka, K. P. Seshadreesan, C. You C, S. Izumi, and J. P.
Dowling, Fundamental precision limit of a Mach-Zehnder interferometric
sensor when one of the inputs is the vacuum, Phys. Rev. A \textbf{96},
052118 (2017).

\bibitem{29} Q. K. Gong, D. Li, C. H. Yuan, Z. Y. Ou, and W. P. Zhang, Phase
estimation of phase shifts in two arms for an SU(1,1) interferomemeter with
coherent and squuezed vacuum states, Chin. Phys. B \textbf{26}, 094205
(2017).

\bibitem{30} C. You, S. Adhikari, X. P. Ma, M. Sasaki, M. Takeoka, and J. P.
Dowling, Conclusive precision bounds for SU(1,1) interferometers, Phy. Rev.
A \textbf{99}, 042122 (2019).

\bibitem{31} G. S. Agarwal, \textit{Quantum Optics} (Cambridge University
Press, Cambridge, 2013).

\bibitem{32} A. Fujiwara and H. Imai, A fibre bundle over manifolds of
quantum channels and its application to quantum statistics, J. Phys. A:
Math. Theor. \textbf{41}, 255304 (2008).

\bibitem{33} R. Demkowicz-Dobrza\'{n}ski, M. Jarzyna, and J. Ko\l ody\'{n}%
ski, \textit{Progress in Optics} (Elsevier, Amsterdam, 2015), Vol. 60 pp.
345--435.

\bibitem{34} C. W. Helstrom, \textit{Quantum Detection and Estimation Theory}
(Academic Press, New York, 1976);

\bibitem{35} S. L. Braunstein and C. M. Caves, Statistical distance and the
geometry of quantum states, Phys. Rev. Lett. \textbf{72}, 3439 (1994).

\bibitem{36} S. Wang and J.D. Zhang, SU(1,1) interferometry with parity
measurement, arXiv:2104.09718, 2021.

\bibitem{37} W. H. Louisell, \textit{Quantum Statistical Properties of
Radiation} (Wiley, New York, 1973).

\bibitem{38} H. Y. Fan, H. L. Lu, and Y. Fan, Newton--Leibniz integration
for ket--bra operators in quantum mechanics and derivation of entangled
state representations, Ann. Phys. \textbf{321}, 480 (2006).

\bibitem{39} R. R. Puri, \textit{Mathematical Methods of Quantum Optics}
(Springer-Verlag, Berlin, 2001).

\bibitem{40} B. M Project, H Bateman, and A Erd\'{e}lyi, \textit{Higher
Transcendental Functions} (McGraw-Hill, New York, 1953).
\end{thebibliography}
\end{document}